\documentclass[aoas,preprint]{imsart}

\RequirePackage[OT1]{fontenc}
\RequirePackage{amsthm,amsmath}
\RequirePackage{natbib}
\RequirePackage[colorlinks,citecolor=blue,urlcolor=blue]{hyperref}

\startlocaldefs
\numberwithin{equation}{section}
\theoremstyle{plain}

\endlocaldefs

\usepackage{amsmath, geometry, amssymb, fullpage, graphicx, amsthm,ifthen,bbm}

\newcommand{\R}{\ensuremath{\mathbb{R}}}

\newcommand{\E}{\mathbb{E}}

\newcommand{\Ac}{\mathcal{A}}
\newcommand{\Bc}{\mathcal{B}}
\newcommand{\Pc}{\mathcal{P}}
\newcommand{\eps}{\varepsilon}

\DeclareMathOperator{\Cov}{Cov}

\begin{document}

\begin{frontmatter}
\title{False Variable Selection Rates in Regression}
\runtitle{False Variable Selection Rates in Regression}

\begin{aug}
  \author{\fnms{Max} \snm{Grazier G'Sell}\thanksref{t1}\ead[label=e1]{maxg@stanford.edu}},
  \author{\fnms{Trevor} \snm{Hastie}\thanksref{t2,t3}\ead[label=e2]{hastie@stanford.edu}}
\and
\author{\fnms{Robert}
\snm{Tibshirani}\thanksref{t4,t5}\ead[label=e3]{tibs@stanford.edu}}

\thankstext{t1}{Supported by NSF GRFP Fellowship}
\thankstext{t2}{Supported by NSF Grant DMS-1007719}
\thankstext{t3}{Supported by NIH Grant RO1-EB001988-15}
\thankstext{t4}{Supported by NSF Grant DMS-9971405}
\thankstext{t5}{Supported by NIH Contract N01-HV-28183}

\runauthor{M. Grazier G'Sell et al.}

\affiliation{Stanford University}

\address{M. Grazier G'Sell\\
Department of Statistics\\
Stanford University\\
Stanford, California 94305\\
USA\\
\printead{e1}\\
\\
\\
R. Tibshirani\\
Departments of Health\\
\quad Research \& Policy and
Statistics\\
Stanford University\\
Stanford, California 94305\\
USA\\
\printead{e3}
}

\address{T. Hastie\\
Departments of Statistics and \\
\quad Health Research \& Policy\\
Stanford University\\
Stanford, California 94305\\
USA\\
\printead{e2}\\}

\end{aug}

\begin{abstract}
There has been recent interest in extending the ideas of False Discovery Rates
(FDR) to variable selection in regression settings.  Traditionally the FDR in
these settings has been defined in terms of the coefficients of the full
regression model.  Recent papers have struggled with controlling this quantity
when the predictors are correlated.  This paper shows that this full model
definition of FDR suffers from unintuitive and potentially undesirable behavior
in the presence of correlated predictors.  We propose a new false
selection error criterion, the False Variable Rate (FVR), that avoids these
problems and behaves in a more intuitive
manner.  We discuss the behavior of this criterion and how it
compares with the traditional FDR, as well as presenting guidelines for
determining which is appropriate in a particular setting.  Finally, we present
a simple estimation procedure for FVR in stepwise variable selection.  We
analyze the performance of this estimator and draw connections to recent
estimators in the literature.
\end{abstract}

\end{frontmatter}

\section{Introduction}\label{intro}

Since the introduction of False Discovery Rates (FDR) \citep{bh1995}, the
idea has had a large impact on error control for many statistical problems and
has inspired many further statistical developments
\citep[e.g.][]{sam,efronlargescale,dudoit2007}.  More recently, there has been an
interest in generalizing the ideas from FDR to variable selection in the
regression setting.

\cite{abramovich2006} introduce the idea of FDR in the regression setting as a
criterion for variable selection, and gives results about the asymptotic
minimaxity of this method.  The results focus on the case where the variables
being considered are orthogonal.  Since then, there has been work extending the
idea of FDR in regression to the correlated variable setting.  These works include
\cite{bg2009}, which proposes a generalized FDR-based penalty to guide
variables selection; \cite{vif}, which proposes a procedure for variable
screening in regression that controls an FDR-related quantity under certain
conditions; \cite{meinshausen2010stability} and \cite{samworth2012stability} on
stability selection; and others \citep[e.g.][]{meinshausen2009p, pseudovar}.

In this paper we consider linear models of the form
\begin{equation}\label{eq:model}
  y_i = \beta_0 + x_i^T \beta + \eps_i,\qquad i=1,\dots,n,
\end{equation}
with $x_i \in \R^p$, $y_i \in \R^n$, $\beta \in \R^p$, and $\eps_i$ independent
and identically distributed.

For variable selection, we denote the selected set of variables as a
subset $\Ac\subseteq\{1,\dots,p\}$ of the potential variables.  The number of
false selections, denoted by $V$ (to be defined carefully in Section \ref{whatisfalse}), is then a property of
the set $\Ac$.  Similarly, the proportion of false selections,
$V/\left|\Ac\right|$, is also a property of the set $\Ac$.

The \emph{rate} of false selections is the expected value of that
proportion of false selections, $\E\left(V/\left|\Ac\right|\right)$, where the
expectation is taken over
realizations of data and conditional on the variable selection procedure.  As a
result, false discovery rates are not a property of a particular selected set $\Ac$, but of the variable selection procedure and
the structure of the model.  In the following sections, we refer to proportions and rates of false
selections in this way.  

In the literature, a false selection in the regression setting is usually
defined as a selected variable that has a zero coefficient in the full model
\citep[e.g.][]{vif, meinshausen2010stability, meinshausen2009p, pseudovar}.
That is, for a set of selected variables $\Ac \subseteq \{1,\dots,p\}$ and 
full model \ref{eq:model}, the proportion of
falsely selected variables is defined as 
\begin{align*}
  \mathrm{FDP} &= \left|\left\{j\in\Ac: \beta_j= 0\right\}\right|/\left|\Ac\right|.
\end{align*}
The $\mathrm{FDR}$ is the expectation of this quantity for the given
procedure.  For this paper, we refer to this as the \emph{full model
definition} of a false selection and FDR.  In contrast to this definition, in Section
\ref{whatisfalse:projection} we introduce new quantities, the False Variable
Proportion (FVP) and its expectation, the False Variable Rate (FVR), which are
defined in terms of the projection of the full model onto the selected
variables $\Ac$. 

The FVR quantity we introduce is motivated by practical issues that arise when
applying FDR to large screening problem such as gene expression studies.  
In these settings, the presence of correlated predictors can lead univariate FDR methods
to select variables that are all
capturing the same underlying signal.  The desire to select unique variables
and to detect multivariate effects has led to the use of regression variable
selection techniques \citep{broman2002model}.  When the full model definition of FDR is
applied in these settings, it is difficult to distinguish the significance of
the correlated predictors, which can inflate their FDR.   This has led to an exploration of other approaches to defining
false selections in the correlated variable setting \citep{frommlet2012}.

In this paper we demonstrate shortcomings of existing definitions of false
selection rates in regression, and propose a new error criterion, called the
False Variable Rate (FVR), which we show exhibits more desirable behavior.
In Section
\ref{whatisfalse}, we discuss what constitutes a falsely select variable,
introduce our new criterion and
examine the differences in these definitions when the predictors
are correlated.  In Section \ref{critdiscuss}, we present intuition for the differences in behavior
of the error criteria, and provide more concrete examples where our
new criteria behaves desirably.  Finally, in Section
\ref{estimation}, we present a simple estimation algorithm for FVR
in stepwise regression.  We provide motivation for this estimator and 
examine the regimes in which it breaks down, making connections to broader
issues in FVR estimation.

\section{Defining a false selection}\label{whatisfalse}

We are interested in examining the population definition of a falsely selected
variable in our regression model.  We begin with a toy example of our problem setting, and then move
on to a discussion of different definitions of a false selection

\subsection{Toy Example}\label{toy}

This toy example may be helpful for gaining an intuition of the alternative
false selection definitions, and for understanding the general purpose of these
criteria.  

\begin{figure}[ht]
  \centering
  \includegraphics[width=0.9\linewidth]{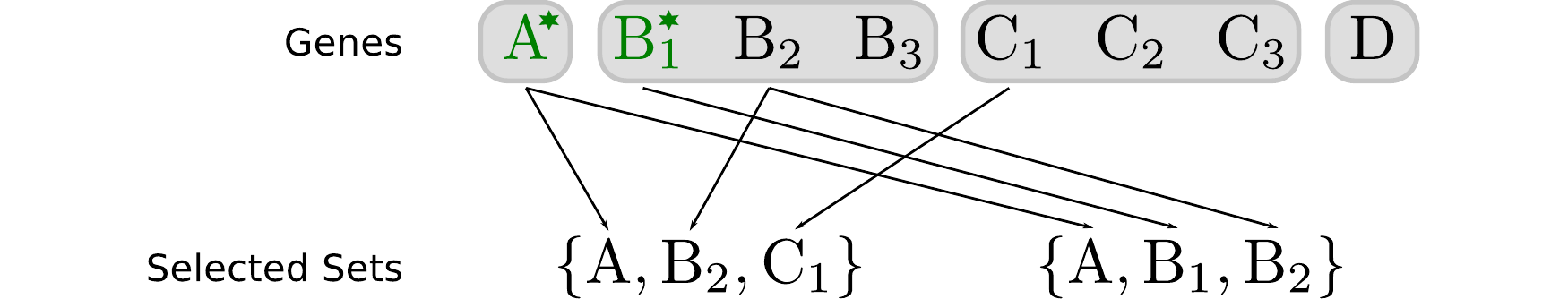}
  \caption{Illustration of a simplified variable selection example.  The eight
  genes represent variables available in our data set.  Genes $A$ and $B_1$
  (green and starred) are biologically relevant to the outcome of interest.  The blocks
  indicate groups of highly correlated variables.  From this data set, we
  consider two possible selected sets of variables.  This paper addresses the
  meaning of false selections in sets like these.}
  \label{img:exsets1}
\end{figure}

Imagine analyzing gene expression data and trying to understand some biological
outcome as a function of that expression.  In our simplified example, illustrated in
Figure \ref{img:exsets1}, we observe the expression of eight genes, $A, B_1, B_2, B_3, C_1,
C_2, C_3, D$.  Of these, genes $A$ and $B_1$ are biologically responsible for
our outcome of interest, and they have corresponding nonzero coefficients in
the full model that contains all the variables.

However, suppose our data have further structure.  Some of the genes occur in a common
biological pathway, leading them to be very strongly correlated.  These groups
are $B_1,B_2,B_3$ and $C_1,C_2,C_3$, illustrated by gray outlines in Figure
\ref{img:exsets1}.  As a result, all the genes from one of these groups carry nearly the same
information about the outcome and are very hard to distinguish based on
experimental data.

The figure illustrates two possible selected subsets of variables from the data
set.  Our goal is to understand how to assess the quality of selected sets like these by
determining a meaningful sense of a correct or false selection, leading to
false selection proportions and rates.  As we see in the following
sections, the presence of correlated predictors allows for different
interpretations of a false selection.  We refer back to this toy example
to help convey the scientific implications of those definitions and
interpretations.

\subsection{Definitions}\label{whatisfalse:definitions}

In these sections, we describe three natural
definitions of a false selection, two of which are common in the literature
and the last which is newly proposed in this paper.

\emph{Note:}  Because this section is focusing on the population definition of
a false selection for a set of variables, rather than for a particular
procedure, we focus on numbers ($V$) and proportions ($\mathrm{FDP}$) of false
selections, rather than rates ($\mathrm{FDR}$).

\subsubsection{Marginal Correlations}\label{whatisfalse:marginal}

The simplest definition of a false selection in our model is similar to the
usual univariate approach to screening using marginal correlations.  It defines
the $j^{th}$ variable to be falsely selected if
$\Cov(y^Tx_j) = 0$.  For a selected set of variables $\Ac$, the number of false
selections $V$ and the false discovery proportion $\mathrm{FDP}$ are given by  
\begin{align}
  V&= \left|\left\{j \in \Ac: \Cov(y,x_j)=0\right\}\right|,\label{eqn:marginalV}\\
    \mathrm{FDP} &= \left|\left\{j \in \Ac: \Cov(y,x_j)=0\right\}\right|/\left|\Ac\right|.\label{eqn:marginalP}
\end{align}

\begin{figure}[ht]
  \centering
  \includegraphics[width=0.9\linewidth]{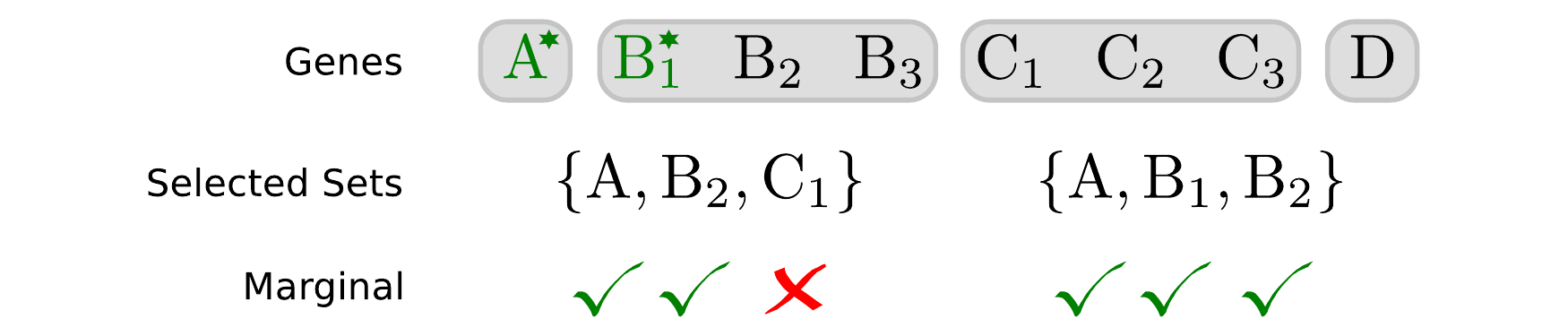}
  \caption{Here we illustrate the marginal definition of a false selection in
  the context of our earlier example (Figure \ref{img:exsets1}).  We see that
  any variable that is marginally correlated with the outcome is considered correct.
  This includes $B_2$ in both sets, since it is correlated with $B_1$ which is in turn
  correlated with the outcome of interest.}
  \label{img:exsets_marg}
\end{figure}

This definition considers a selected variable interesting if that variable
captures information about the signal on its own, irrespective of any of the
other variables in the data set or in the selected model.  In our gene
expression example, any pathway with a gene that is important to the outcome
will have all of the genes in that pathway considered correct selections, since
they will all have marginal correlation with the outcome (Figure
\ref{img:exsets_marg}).

This definition of a false selection is equivalent to the one used in many variable selection
screening procedures, for example \cite{sam} and \cite{efronlargescale}.  We will not
focus much on controlling this false selection rate, since it can be handled
with the standard univariate FDR tools that have been very well discussed in
the literature.

\subsubsection{Full Model Definition}\label{whatisfalse:full}
As mentioned in the Section \ref{intro}, the literature usually defines a falsely
selected variable as one which has a zero coefficient in the full model.  In
the notation from Equation \ref{eq:model}, this means that $V$ and
$\mathrm{FDP}$ are given by
\begin{align}
  V &= \left|\left\{j\in\Ac: \beta_j= 0\right\}\right|\label{eqn:fullV}\\
  \mathrm{FDP} &= \left|\left\{j\in\Ac: \beta_j=
  0\right\}\right|/\left|\Ac\right|.\label{eqn:fullP}
\end{align}
This definition has been used in several papers, among them \cite{meinshausen2010stability,
meinshausen2009p, pseudovar}. A
modified version of the FDR appears in \cite{vif}, using this definition of the number of false discoveries $V$.

This definition of a false discovery is natural, particularly in the setting
with uncorrelated $x_j$.  In that setting, it actually agrees with the
definition in Equations (\ref{eqn:marginalV}, \ref{eqn:marginalP}).  When the $x_j$ are
correlated, the meanings of these two definitions differ.  The definition in
terms of marginal correlations, as we mentioned, asks if each selected variable
captures information about the signal on its own.

\begin{figure}[ht]
  \centering
  \includegraphics[width=0.9\linewidth]{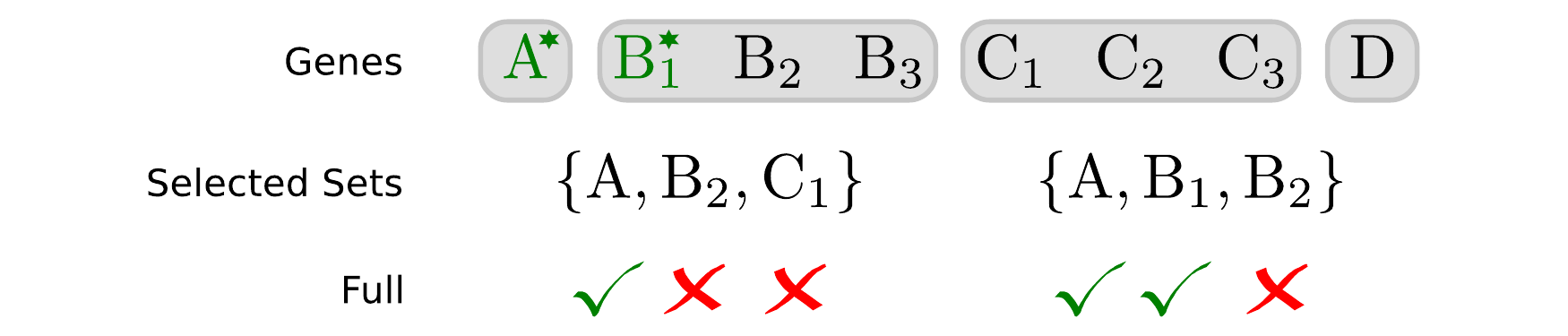}
  \caption{Here we illustrate the full model definition of a false selection in
  the context of Figure \ref{img:exsets1}.  Here only variables $A$ and $B_1$
  can be considered correct detections, since they are the only variables with
  nonzero coefficients in the full model.}
  \label{img:exsets_full}
\end{figure}

In contrast, the coefficient
corresponding to a variable in the full model is only nonzero if that variable
captures information about the signal that is not captured by any other
variables in the full model.  The proportion of false selections in this
setting therefore corresponds to the fraction of selected model variables that
fail to uniquely capture signal among all the variables in the considered data
set.  In our gene expression example, only the gene from an important pathway
that has nonzero coefficient in the full model will be considered important (in
our case $A$ or $B_1$, see Figure \ref{img:exsets_full}).  This can be
counterintuitive, since any one of the genes from these pathways could be strongly predictive of
the outcome of interest.

Furthermore, in practice it would likely
be impossible to determine which of the highly correlated variables in an
important pathway actually carried the nonzero coefficient in the full model,
which would lead to all the variables appearing as false selections according to
this criterion.

This is a reasonable choice for some statistical problems and some scientific
settings.  It has a strong connection to the full model $p$-values in regression, where significance of a coefficient shows that
a particular variable is significantly correlated with the response after
the effects of all the other variables have been removed.  However, as we 
discuss further in Section \ref{critdiscuss}, this criterion has unintuitive
behavior for many of the scientific problems in which variable selection is
being applied.  In the next section, we introduce a new criterion which
is more appropriate for those settings.

\subsubsection{A new approach: False Variable Rate}\label{whatisfalse:projection}
In this section, we propose a new approach for defining false selections which
lies between these two extremes.  Rather than requiring that an interesting variable be
correlated with the signal in a way that is not explained by any other
variables in the data set, we instead consider a variable to be an interesting
selection if it captures signal that has not been explained by any other
variable \emph{in the selected model}.

For many of the situations where variable selection is applied, this is a more
natural view.  Common variable selection approaches like stepwise or
$L_1$ regression attempt to include variables that capture part of the signal
that the other selected variables miss.  However, neither of these methods
check that a variable being included captures signal that \emph{excluded}
variables do not also capture.  For applications like the screening of
predictors in biology, where predictors may be strongly correlated and the
data matrix may not be carefully structured with meaningfully chosen columns, this is a more interpretable criterion.  We come back to
this idea when we contrast the methods more carefully in Section
\ref{critdiscuss}.

We now define a criterion with the desired behavior.  Rather than looking at
the coefficients of the full model as in Section \ref{whatisfalse:full}, we
instead look at the coefficients of the model formed by projecting the true
model onto the selected variables.  This resembles some ideas from \cite{posi},
where inference is conducted with respect to the selected model even if there
may be some larger true model.

For a selected set $\Ac\subseteq\{1,\dots,p\}$, we have a restricted
data matrix $X_{\Ac}$, formed by the columns with indices in $\Ac$.  We can
project the mean $X\beta$ from the full linear model onto this subset of predictors
$X_{\Ac}$.  This gives a projected mean $X_{\Ac}\beta^{(\Ac)}$, for some $\beta
\in \R^{|\Ac|}$.  In the event that $\beta^{(\Ac)}$ of this form is not unique,
meaning that $X_{\Ac}$ is not full rank, we choose $\beta^{(\Ac)}$ to be any of
the sparsest vectors satisfying the projection requirement.

We now define a selected variable to be a false selection if it has a zero
coefficient in this vector.  This means that the number of false selections is
just the number of zeros in this vector of coefficients for the projected mean,
giving
\begin{align}
  \mathrm{V} &= \left|\left\{j\in\Ac: \beta_j^{(\Ac)} =
  0\right\}\right|\label{eqn:projV}\\
  \mathrm{FVP} &= \left|\left\{j\in\Ac: \beta_j^{(\Ac)} =
  0\right\}\right|/\left|\Ac\right|\label{eqn:projP}
\end{align}
where we refer to the proportion of falsely selected variables by this
definition as the false variable proportion (FVP) to differentiate it from the
usual regression definition (FDP) in Section \ref{whatisfalse:full}.  Similarly, the
expectation of this proportion is referred to as the False Variable Rate, or
FVR.

\begin{figure}[ht]
  \centering
  \includegraphics[width=0.9\linewidth]{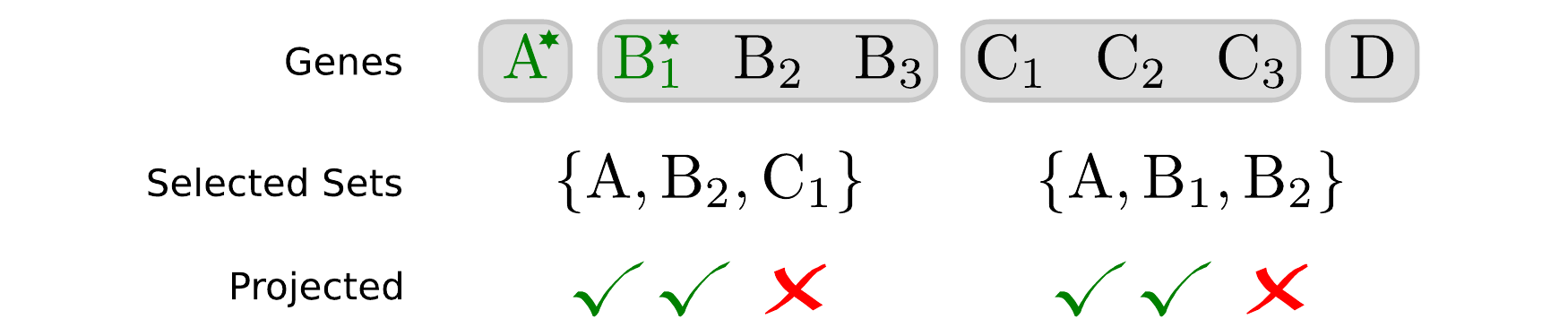}
  \caption{Here we illustrate our new projected model definition of a false selection in
  the context of Figure \ref{img:exsets1}.  We see that variables are now
  correct selections if they are capturing unique signal among the selected
  variables.  Thus $B_2$ is correctly selected in the first set.  However,
  $B_2$ is considered a false selection in the second set because it adds no
  information beyond $B_1$.}
  \label{img:exsets_proj}
\end{figure}

This quantity has the desired interpretation.  A variable is considered interesting if it
is correlated with the signal $y$ after the effects of the other \emph{selected} variables
have been removed.  If two variables are capturing the same piece of signal,
including either one of them will be a good selection, but including a second
one will not be adding any new information, and will thus be a false selection.

In our gene expression example, this means that one gene selected from a
given influential pathway will be considered a correct selection and any
further selections from that pathway will be incorrect.  This is illustrated in
Figure \ref{img:exsets_proj}, where we see that the selection of $B_2$ in the
first set is considered correct, because it is adding information to the
selected set.  In the second set, $B_2$ is considered incorrect, because $B_1$
is already contributing the same information about the outcome.  
This seems like a natural definition in this setting.  In Section \ref{critdiscuss}, we elaborate on the differences between the
criteria in detail.

\emph{Note:} Some care needs to be taken for models with random $X$, where we
want to rule out spurious correlations between the random predictors.  The
number of correct selections in $\Ac$ is generalized to be the size of the
smallest subset $\Bc\subseteq \Ac$ such that, when conditioning on $\Bc$, $y$
is conditionally uncorrelated with the rest of $\Ac\backslash \Bc$.  This has a
nice form for Gaussian graphical models, discussed in Section
\ref{graphicalmodel}.

\subsection{Summary and comparison}\label{whatisfalse:summary}
\begin{figure}[ht]
  \centering
  \includegraphics[width=0.9\linewidth]{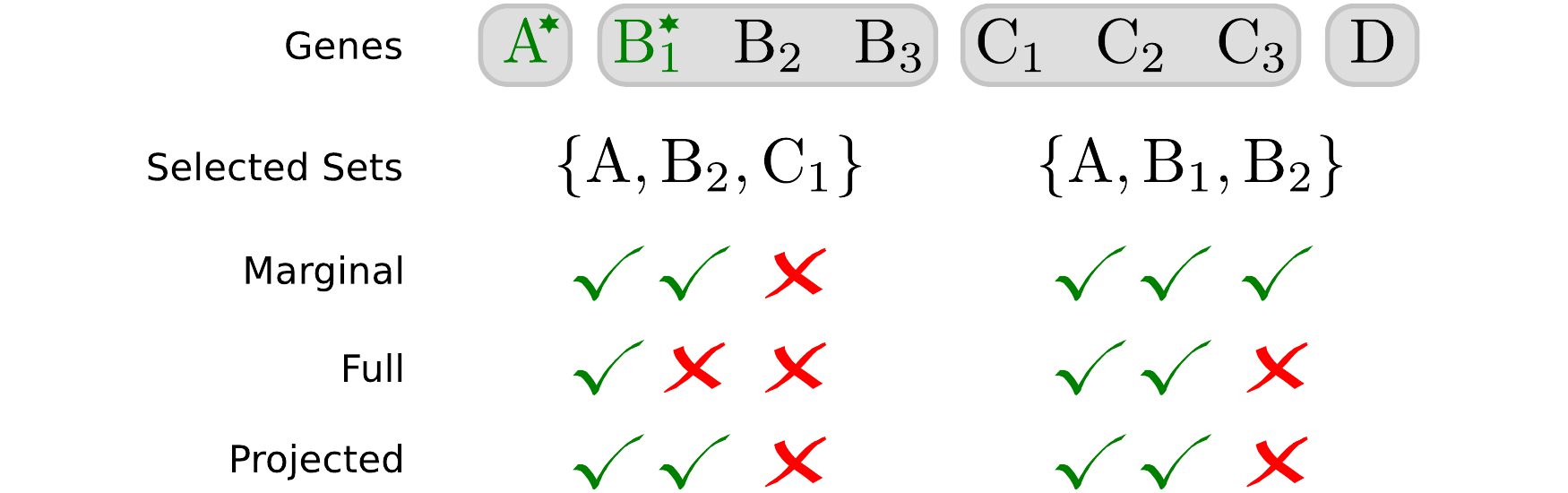}
  \caption{A summary of the implications of the marginal, full, and projected
  model definitions of a false selection on our example from Figure
  \ref{img:exsets1}.  We see that there are cases in which each of the
  definitions disagree with the others.}
  \label{img:exsets_all}
\end{figure}

Before moving on to discussion of implications and behavior of the different
error criteria, we briefly summarize the three definitions that we have discussed,
and their simple interpretation.  Their implications for the gene expression
example of Section \ref{toy} are shown in Figure \ref{img:exsets_all}.

\emph{Marginal view (Section \ref{whatisfalse:marginal}).}  A variable $x_j$ is
considered a false selection if $\Cov(y,x_j) = 0$.  This implies that a variable
is interesting if it is correlated with the signal, without regard to any of
the other variables in the data set.

\emph{Full Model view (Section \ref{whatisfalse:full}).}  A variable $x_j$ is
considered a false selection if $\beta_j = 0$ in the full model.  This implies
that a variable is interesting if it is correlated with the signal
after conditioning on all the other variables \emph{in the entire data set}.

\emph{Projected Model view (Section \ref{whatisfalse:projection}).}  A variable $x_j$
is considered a false selection if $\beta_j^{(\Ac)} = 0$ in the projected model
onto the selected variables $X_{\Ac}$.  This implies that a variable is an
interesting selection if it is correlated with the signal after conditioning on all the other variables \emph{in the set of selected
variables}.  This definition is used for our proposed False Variable Rate (FVR)
criterion.

In the next section, we will see the impact of the differences in these
definitions on the behaviors of the error criteria.

\section{Contrasting the False Selection Criteria}\label{critdiscuss}

In this section, we will discuss the behavior of the usual full model
definition and our new projected model definition of false variable
selection rates.  We will see that, though the full model definition is
reasonable in some settings, it leads to non-intuitive behavior in
common variable selection settings. We will show that our proposed approach
has intuitive and desirable behavior in those cases.

\subsection{A simple example}\label{critdiscuss:example}

\begin{figure}[ht]
  \centering
  \includegraphics[width=.9\linewidth]{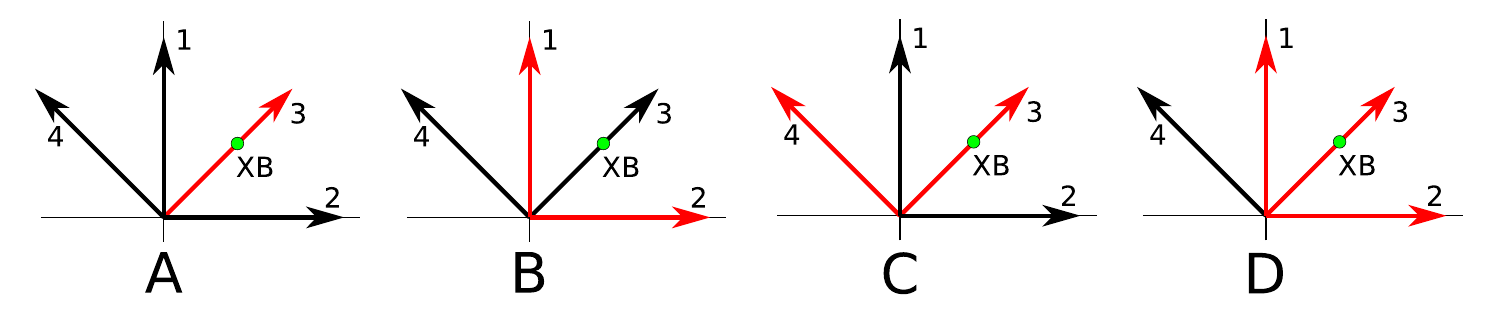}  \caption{Representation
  of four possible variables.  The projection of the true (noiseless) model
  into the space spanned by the variables is indicated as
  the green circle and label $X\beta$.  Variable 3 is perfectly
  correlated with the signal, variables 1 and 2 are correlated with the signal,
  and variable 4 is orthogonal to the signal.  The red arrows indicate the
  variables that have been selected.}\label{img:example}
\end{figure}

To understand the differences in behavior between these definitions, we begin
with the following
simple example, represented in Figure \ref{img:example}

\begin{table}[tbh]
  \centering
  \begin{tabular}{r|c|c|c|}
    & Marginal FDP & Full Model FDP & Proposed FVP\\
    \hline
    Figure A & 0 & 0 & 0 \\
    Figure B & 0 & 1 & 0 \\
    Figure C & 1/2 & 1/2 & 1/2 \\
    Figure D & 0 & 2/3 & 2/3\\
    \hline
  \end{tabular}
  \caption{Resulting false selection proportions from applying each of the
  three definitions of Section \ref{whatisfalse:definitions} to the scenarios in Figure
  \ref{img:example}.  The criteria disagree on values in Figures B and D
  because the definitions give different value to correlated variables that
  explain the same part of the signal.}
  \label{fdrexampletable}
\end{table}

We consider four different simple cases here.  Note that we are examining
proportions of false selections for a selected set, rather than rates of false
selections for a procedure, so we will describe FDP/FVP quantities, rather than
FDR/FVR rates.  The false selection proportions for each scenario and each
definition are shown in Table \ref{fdrexampletable}.  

All three definitions agree on scenarios A and C, since they deal only with
variables that are perfectly correlated or orthogonal to the true signal.  The
cases B and D are more interesting.   In case B, variables 1 and 2 are
correct selections according to the projected definition, since they capture
information about the signal that is not included in any other \emph{selected}
variable.  These variables are both considered false selections by the full
model definition, since the data set (though not the selected set) contains
variable 3, which captures all the information in variables 1 and 2.

In scenario D, the full and projected model definitions now agree, since
variable 3 is included in the selected set, rendering variables 1 and 2
uninformative.  The marginal definition continues to consider all three
variables correct selections, since it is not concerned with uniqueness.

We see that the definitions may all disagree, depending on the structure of the
data set and the selected model.  In general, the full and projected model
approaches differ when variables are selected that are correlated with the
signal, but would have their correlation explained away by an unselected
variable in the data set.

\subsection{Graphical Model View}\label{graphicalmodel}

\begin{figure}[ht]
  \centering
  \includegraphics[width=.5\linewidth]{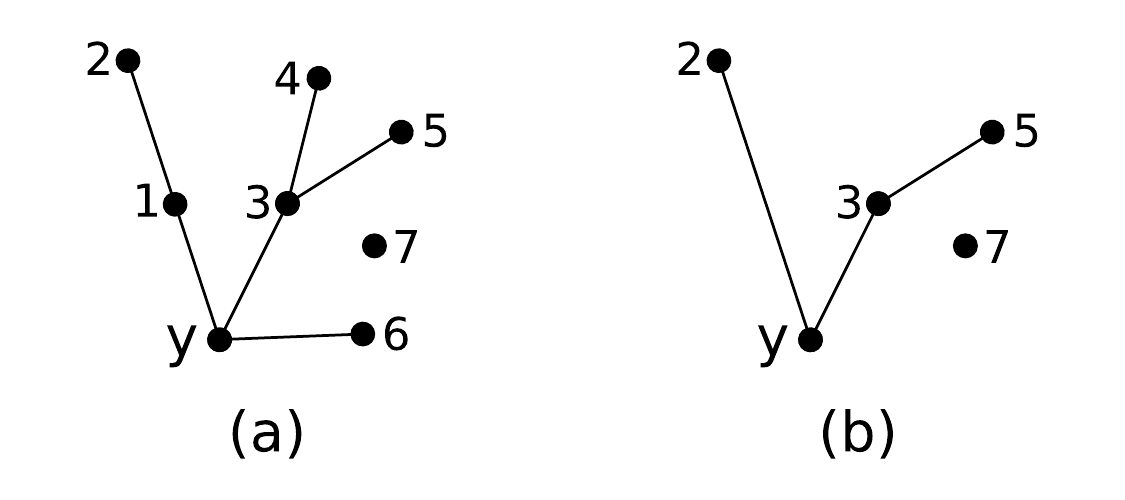}
  \caption{Example of a dependence graph.  Plot (a) shows the dependence graph
  for the full model.  Plot (b) shows the induced dependence graph for the
  marginal distribution of the selected variables $\Ac=\{2,3,5,7\}$ and $Y$.}
  \label{img:graph}
\end{figure}

The interpretation of and differences between these false selection definitions
are particularly clear in a Gaussian graphical model setting.   Suppose that the variables
$X_j$ and the response $Y$ have a joint Gaussian distribution, with
distributions $X
\sim N(0,\Sigma)$ and $Y \sim N(X^T\beta,\sigma^2)$.  We represent the dependence structure
of the variables by the usual dependence graph, as illustrated in
Figure \ref{img:graph}a.  Two variables are connected here if they have nonzero
partial correlation after conditioning on the other variables.

If we select a subset of the variables $\Ac$, there is an induced dependence
graph for the marginal distribution of $y\cup\Ac$.  The structure follows from
the usual manipulations of Gaussian covariance matrices (see Appendix
\ref{app:details:gaussian}).  This marginal graph
corresponds to the dependence structure of the variables in the projected model we discussed
earlier.

Each of our definitions of a false selection have an interpretation in terms of
these graphical models.  Suppose we select variables $\Ac = \{2,3,5,7\}$ in the
example shown in Figure \ref{img:graph}.  In the marginal definition, a
variable in $\Ac$ is a false selection if it does not have a path to $y$ in the
full graph $(a)$, since such a path would correspond to a marginal correlation
with $y$.  In this case, all the variables are connected to $y$ except
variable 7, so
there would be one false selection by that definition, and $\mathrm{FDP} = 1/4$.

In the full model definition, a variable in $\Ac$ is a false selection if it is not
directly connected to $y$ in the full graph $(a)$, since such links correspond
to nonzero partial correlations.  By this definition,
variables 2,5,7 would be false selections, and the $\mathrm{FDP} = 3/4$ for
this approach.

In our projected model definition, a variable in $\Ac$ is a false selection if
it is not directly connected to $y$ in the graph $(b)$ induced on the selected
variables.  The links in that induced graph correspond to correlations that
cannot be explained by other selected variables.  In the example, variables 2 and 3 are directly connected to $y$ in
the induced graph, and variables 5 and 7 are not, so the $\mathrm{FVP} = 1/2$.

From these graphs, we gain an intuition for the behavior of the full and
projected definitions.  By switching from the full model to the projected
model, represented by the induced graph, unselected variables that were
important in the full model can induce importance in correlated variables.  
This lets
selected variables that are carrying that missing information be considered
correct selections when they are included. 

\subsection{Implications for the Choice of False Selection Criteria}\label{implications}

So far in this section, we have described the differences between the false
selection criteria we have presented.  In this section, we will discuss the practical implications of
these differences in statistical problems.  We show that in many common
settings the usual full model FDP/FDR will give misleading results, and that
our proposed FVP/FVR definition will have the intuitive behavior we desire.  

\subsubsection{Highly correlated predictors}\label{implications:correlated}

Suppose our dataset contains highly
correlated predictors.  This commonly occurs in biological data, where there
might be an underlying factor driving several variables.  In our gene
expression example, several genes in our
data set could be from the same biological pathway, leading them to be
over-expressed or under-expressed together.

Now imagine that one of these pathways is biologically relevant to our outcome
of interest.  This could lead to several variables which are strongly correlated
with the outcome, but are also highly correlated with one another.  In the
setup we just described, it may not be possible to distinguish which of these variables
has the ``true'' relationship with the signal variable.  In that case, what behavior do we
desire from an error criterion and how do the criteria we have
proposed behave?

For the usual univariate definition of a false discovery, all of these
correlated variables will be considered correct detections, since they carry
some information about the variable.  However, this is not likely to be the
behavior we are interested in, since the use of a multivariate model in the
first place expresses interest in capturing ``unique'' signal of some
kind. 

In the full model definition of a false discovery, each of these
variables is ``interesting'' only if they capture \emph{unique} signal among
all the variables being considered.  Because of the high collinearity, it will
be impossible to distinguish a unique signal that is captured by any one of the
variables.  As a result, they would all appear as false selections in
practice.  

Furthermore, if the selection procedure were guided by an attempt to
control the full model FDR, it would discourage the selection of any of these
variables because of the other highly correlated variables in the data set.  As
a result, it would be likely that such a procedure would fail to select any of
these variables, even though any one of the variables could carry most of the predictive power of the data set.

Contrast these behaviors with that of the proposed FVR criterion.
As we have discussed, a variable is considered interesting by FVR if it
captures unique signal among all the \emph{selected} variables.  Consider a
selected set that contains only one of the highly correlated and predictive
variables.  While the full model definition would consider it a false
detection, the FVR definition considers it a correct selection because it is adding
unique explanatory power to the \emph{selected} set of variables.  Furthermore, if several of these correlated
explanatory variables are added, only one of them (it does not matter which)
will be considered a true detection by FVR and the others will be considered false.
This makes intuitive sense, as the group of correlated variables should be able
to contribute ``one variable'' of explanatory power.

We believe that in many settings, this is the interpretation that is being
sought
for a false discovery rate.  It is helpful to have a criterion that agrees with
this interpretation, and it is
convenient that defining the criteria in terms of the projected model gives
this interpretation.

\subsubsection{Stability to the set of considered variables}
\begin{figure}[ht]
  \centering
  \includegraphics{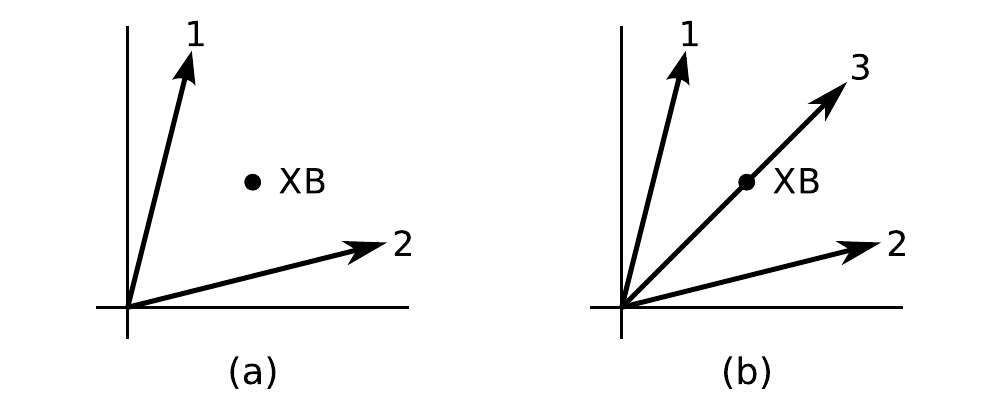}
  \caption{Simplified example demonstrating full model FDR sensitivity to
  excluded variables.  In scenario (a), both 1 and 2 would be considered
  correct selections.  In scenario (b), both 1 and 2 are considered false
  selections, whether or not variable 3 is selected.}
  \label{img:threevar}
\end{figure}

Another non-intuitive property of the full model FDR concerns its stability to
changes in the data set.  Suppose that our data
set contains just two variables (call them 1 and 2), each of which captures some of the signal in the
dependent variable.  This scenario is shown in Figure \ref{img:threevar}(a).
Imagine that a variable selection procedure selects both of these variables.  
Under the full model definition of FDR, both of these selections are considered
correct.  Now suppose that the data set had included another variable 3 that
is very well correlated with the signal, so that variables 1 and 2 are
uncorrelated with $y$ conditional on variable 3 (Figure \ref{img:threevar}(b)).  Then under the
full model definition of FDR, both variables 1 and 2 would now be considered
false detections, \emph{even if variable 3 was not selected}!  The FDP of the
selected set changes from 0 to 1, without any change to the variables included
in that selected set.

This emphasizes
that the full model FDP is not a property of a selected set, but of both a
selected set and of the full data set being considered.  The FDP is incredibly
unstable to changes in the full set of variables being considered.  The
implications of this are worrisome in settings like gene screening, where the
variables being measured may be a matter of convenience and the existing
microarrays, rather than careful design for a particular experiment.  Meaningful
interpretation of the full model FDR depends very heavily on an understanding
of the entire data sample that was collected and analyzed.

\subsubsection{Summary}

In this section we described practical examples of how the full model
definition of FDR
clashes with the intuitive interpretation of a false discovery rate when the
predictors are correlated.  We saw that when several correlated variables are
capturing essentially the same signal, the FDP can be unstable and, in the
presence of noise, the FDR for the selected variables can be high even if
the variables are capturing a strongly predictive signal.  In contrast, the
FVP and FVR behave intuitively in the presence of correlated explanatory
variables such as these, counting the first such variable to be selected as an interesting
selection, and any of the following such variables uninformative.

In addition, we showed that the FDP and FDR are highly sensitive to all the
variables included in the data set, even those variables that
are not selected.  As a result, the usefulness of the full model FDR depends on a careful
understanding of all the variables being considered in the data set.  The FVR
avoids some of this trouble, since it is not affected by unselected variables
in the data set.

It is worth noting that there are cases where the full model FDR is the correct
definition to use.  First, if the predictors are uncorrelated, all three definitions of a false selection are equivalent.  Second,
there are scientific setups where the interpretations described above are the
desired ones.  Suppose one has carefully selected all the variables being
considered, and that a guarantee is desired that any selected variable is
actually uniquely related to the signal among all the variables being
considered.  Then one might desire to only select a variable if its relation
without outcome $y$ stands out among all the variables.  If two variables
cannot be distinguished, the scientist might not wish to select either until
enough data can be gathered to distinguish between them.  In that setting, the FDR
definition will have the proper interpretation.  However, we feel that in the
majority of experimental setups, particularly experiments where the focus is
on screening, the FVR definition is more in line with research
goals and intuitive interpretations.

We conclude this section with a simple simulation.  This simulation
demonstrates the differences in behavior between FDR and FVR when evaluating
stepwise regression on a set of correlated variables.  The simulation is
constructed to fit the setting of Section \ref{implications:correlated}, where
predictive variables are present but appear in strongly
correlated groups.

\subsection{Simulation Example}\label{understanding:simulation}

\begin{figure}[ht]
  \centering
  \includegraphics[width=.8\linewidth]{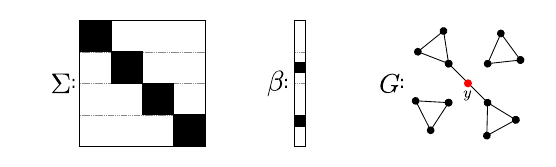}
  \caption{Illustration of the simulation setup for Section
  \ref{understanding:simulation}.  Nonzero elements of $\Sigma$ and $\beta$ are
  shown by the shaded rectangles, and the graph illustrates the corresponding
  joint dependence structure of $X$ and $y$.}
  \label{img:matrixpic}
\end{figure}

To see the difference in behavior between the full model definition and our
projected model definition, consider the following simple setup, illustrated
in Figure \ref{img:matrixpic}.  

We create several blocks of variables, each of which is highly correlated
internally.  For a subset of the blocks, we select one variable within the block to
have a nonzero coefficient in $\beta$.  The structures of the resulting matrices are shown in
\ref{img:matrixpic}, along with the corresponding dependence graph.

We choose this setup because we expect it to demonstrate a strong difference
between the methods.  The correlation within the blocks is strong enough that
the selection method will be unable to distinguish the true signal variable
within each correlated block, but the signal associated with the nonzero coefficients
is large enough to be detected at the group level.  As a result, selection methods should be able
to pick those blocks that have signal, but choosing the ``correct'' variable
within each block should be nearly random.  By the usual full model definition
of a false selection, these selections will be counted as false.  By our new
definition, it is recognized that any variable within that block carries
essentially the same signal, so the first selection within each block will be
considered correct.  

\begin{figure}[ht]
  \centering
  \includegraphics[width=.8\linewidth]{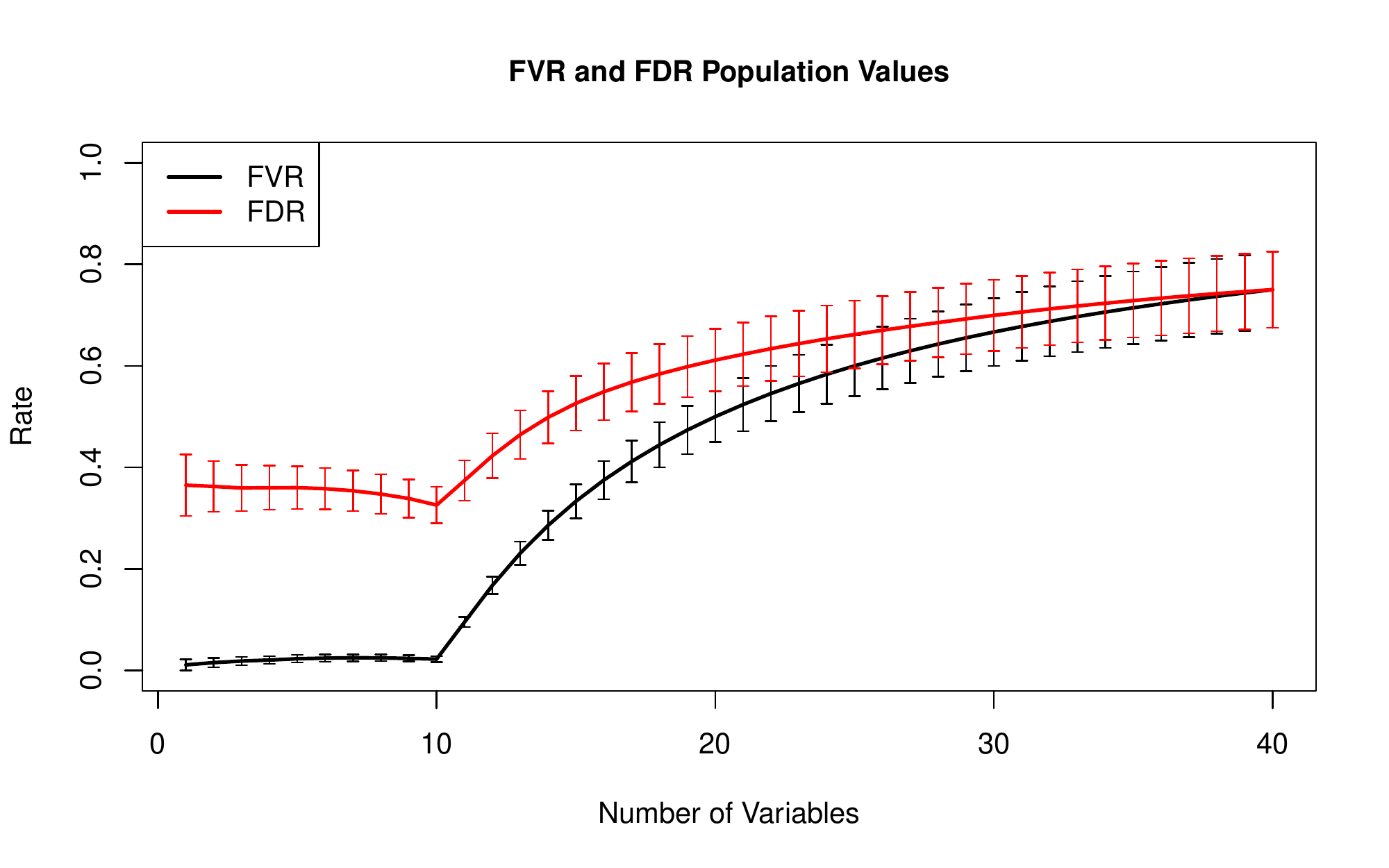}
  \caption{True full model FDR (black) and projected model FVR (red) for the simulation
  in Section \ref{understanding:simulation}.  Stepwise selection is carried out
  on a simulation with 20 pairs of correlated variables, 10 of which contain
  one important signal variable.  We see that FVR considers either variable of an
  important pair correct, while FDR does not.  Monte Carlo standard errors are
  shown.}
  \label{img:fvrplot}
\end{figure}

The plots of the true population FDR and FVR are shown in Figure
\ref{img:fvrplot}.  This particular simulation has 20 blocks of two variables
each.  Ten of those blocks are selected to have a nonzero coefficient in
$\beta$ for one of the variables.  The correlation within each block is $0.95$,
the additional
noise on $y$ is $0.8$, and the total number of observations is $n=50$.

We see that the usual FDR criterion finds a false discovery rate of about 40\%
for the first 10 selections, while our new criterion gives an FVR of nearly
zero for the first 10 selections.  These ten selections correspond to correctly
selecting the ten blocks with signal, though not necessarily the variable
within those blocks with nonzero $\beta$ coefficient.

While there are situations where both criteria could make sense, we believe
that in many settings, the value reported here for FVR is more in line with
peoples interpretation and goals in these settings.  The first ten variable
selections were highly predictive about $y$, and it is often reasonable to have
a sense of false discovery rate that is in line with this fact.

\section{Estimation of False Variable Rates}\label{estimation}

In this section, we discuss estimation of FVR.  Classically, estimation of FDR in the regression setting has been
quite difficult when the variables are correlated \citep[e.g.]{vif}.  We
expect the FVR to be easier to control, since it is more closely related to
traditional variable selection procedures.  Nevertheless, more development of
good estimation and control procedures is needed.

In this section, we will construct a very simple, illustrative estimate of FVR
for stepwise regression.
 This is not intended to be the ideal estimator of FVR, it
should instead be viewed as an illustration of FVR and how one might approach its
estimation.  Through simulation, we will demonstrate that this simple
estimator works reasonably well, even with correlated predictors, providing
further evidence that FVR may be an easier target for control.

We
will also examine the assumptions behind this estimation scheme and the regimes in which this simple estimation procedure breaks down.  These
weaknesses are instructive, as our procedure shares those limitations with
several recent methods for controlling false selections in regression.
We hope that a better understanding of this simple estimator can
inform the construction of future estimators.

\subsection{Motivation for the Estimator}\label{estimation:motivation}

The motivation for this algorithm comes from the idea that at each step of stepwise regression, the procedure admits the variable that
appears to capture the most signal \emph{once the effects of the other
currently selected
variables have been removed}.  This approach is in the same spirit as the FVR
criterion we have been discussing. 

At each step, we can imagine the hypothesis that the new
model is a true improvement over the old.  This is the usual statistical
hypothesis
comparing two nested models, where we test whether the coefficient of the new
variable is zero in the larger of the two models.  We will neglect for a moment that
the selection of the variable would make the usual tests for this hypothesis
invalid.

If we look at the number of these incremental hypotheses that are ``null,'' this seems
similar to the number of false selections as defined in the FVR.  While this
statement is not necessarily true in reality, there is a relationship between these two
quantities.  We will make this relationship more explicit later in this
section and justify it more carefully.  Using this relationship, our
simple algorithm will amount to estimating the number of null incremental
hypotheses that were traversed in arriving at our selected model.

\subsection{Algorithm}\label{estimation:algorithm}

Before delving into a more careful justification of our algorithm, we present
it in its complete form here.

To avoid issues with inference after selection, this algorithm relies on random
splits of the data to separate the selection and
inference stages.  The following algorithm describes the action for one random
split of the data.
\begin{enumerate}
  \item Split the data randomly into two pieces, call them $X^{(1)}$ and
    $X^{(2)}$.  For convenience we use even splits, though this is not necessary.
  \item On $X^{(1)}$, fit a stepwise selection path.  This gives an ordering
    $\Pc$ to all $p$ of our variables, by the order in which they are selected.
    Notationally, we define $\Pc_j$ to be the set of the first $j$ variables in
    the ordering.
  \item Define the incremental hypotheses $H_j^{(\Pc)}$ as follows.  Let $\ell_j$
    be the variable added at the $j^{th}$ step in $\Pc$.  Let $\beta^{(\Pc_j)}$ be the coefficients of the
    model projected on $X_{\Pc_j}$, the $j$ variables selected by step $j$.  Then
    \begin{align*}
      H_j^{(\Pc)} &: \beta_{\ell_j}^{(\Pc_j)} = 0,
    \end{align*}
    which is just the hypothesis that the $j^{th}$ addition was not a useful
    one (at the point that it was added).

    For each hypothesis $H_j^{(\Pc)}$, we can obtain a $p$-value $p_j$ through the
    usual $F$ or $t$ test of the nested models, using the data from $X^{(2)}$.
    This inference is valid because our variables were not selected on
    $X^{(2)}$.  Nonparametric tests, like those based on permutations or the
    bootstrap, can also be used here to avoid distributional assumptions.
  \item For each model size $k$, we now estimate the number of null hypotheses in our set of
    hypotheses $\{H_j^{(\Pc)}; j=1,\dots,k\}$.  We use a threshold estimator as in
    \cite{storey2002}, giving the estimator 
    \begin{align*}
      \hat{V}_k^{(\lambda)} &= \frac{\#\{p_j > \lambda; j\le k\}}{(1-\lambda)}
    \end{align*}
    for any threshold $\lambda \in (0,1)$.  We will show that $\E\hat{V}_k$ bounds
    the number of null hypotheses in $\{H_j^{(\Pc)}; j\le k\}$.
  \item The estimate from this split for the FVR for model size $k$ is
    $\hat{V}_k^{(\lambda)}/k$.
\end{enumerate}

We average these estimates, $\hat{V}_k^{(\lambda)}/k$, over many splits of $X$ to obtain a final estimated
FVR for each selected model size.

In the remainder of this section, we will provide justification for this
algorithm and present simulation results.

\subsection{Justification}\label{justify}

In this section we will provide justification for the estimation procedure
presented in Section \ref{estimation:algorithm}.  Due to the length of some of
these explanations, some of the details have been moved to Appendix
\ref{app:justify} and
summaries are presented here.

There are three pieces of this algorithm that require justification.
\begin{enumerate}
  \item That the hypotheses $H_j^{(\Pc)}$ resulting from the selected path $\Pc$ are
    appropriate hypotheses to be looking at, in the sense that the number of
    null hypotheses in $\{H_j^{(\Pc)}\}$ should correspond to the number of
    false selections appearing in the FVR definition.
  \item That we are estimating the number of true null hypotheses in the set
    $\{H_j^{(\Pc)}\}$ in an appropriate way.
  \item That splitting the data give reasonable estimates of the quantity of
    interest.
\end{enumerate}

We will address each of these points in the following subsections.

\subsubsection{Appropriateness of $H_j^{(\Pc)}$}\label{justify:ordering}

Here we argue that the hypotheses $\{H_j^{(\Pc)}\}$ corresponding to the
steps of the selected path are reasonable hypotheses to consider, in the sense
that the number of true null hypotheses in
$\{H_j^{(\Pc)}\}$ should be a good estimate of the number of variables with zero
coefficients in the final projected model.  In cases where the variables are
correlated, there is no reason that this should be true for the incremental
hypotheses corresponding to a general ordering
of the variables.  

The details of this argument can be found in Appendix \ref{app:justify}.  The
general idea is that there exist particular orderings of the variables for which the
number of null incremental hypotheses is exactly the number of zero
coefficients in the projected model.  Furthermore, the ordering produced by
stepwise selection is not too far from these ideal orderings, so the resulting
estimate is not badly biased. 

This dependence on the ordering of the variables has implications for FVR and
FDR estimation.  One is that the estimation method of Section
\ref{estimation:algorithm} can only be extended to selection methods that provide a reasonable ordering of the variables.
Similarly, one should be wary of potential bias in other methods that estimate FVR or FDR based on the incremental hypotheses along a
variable ordering, particularly those that rely on random or arbitrary
orderings.

\subsubsection{Justification of the threshold estimator}\label{justify:storey}

The next piece of the algorithm uses a threshold estimator, as in
\cite{storey2002}, to
estimate the number of true null hypotheses in $\{H_j^{(\Pc)}\}$ based on the
$p$-values $p_j$. 

We can show that the expectation of the threshold estimator
$\hat{V}_k^{(\lambda)}$ is an upper bound on the number of true null
hypotheses in $\{H_j^{(\Pc)}; j\le k\}$:

\begin{align*}
  \E \hat{V}_k^{(\lambda)} &= \frac{1}{1-\lambda} \sum_{j=1}^k \E I_{p_j >
  \lambda}
  \ge \frac{1}{1-\lambda} \sum_{\stackrel{j=1}{H_j\text{ true null}}}^k \E I_{p_j
  > \lambda}
  = \#\{j\le k: H_j\text{ true null}\}.
\end{align*}

This bound depends only on the $p$-values from the true null hypotheses, so the bound will be loose when there is a large contribution
from the false null hypotheses.  This will happen when the $p_i$
corresponding to the false nulls have a significant probability of exceeding
$\lambda$, or when there are a large
number of false null hypotheses in the selected set.  This is demonstrated in
the simulation in Figure \ref{estsim:high}.

This also suggests a bias-variance trade off when selecting $\lambda$, as a
large $\lambda$ will give a smaller bias but a larger variance.  Appendix \ref{app:justify:threshold}
mentions a bootstrap approach to calibrating $\lambda$, like that of \cite{storey2002}.

\subsubsection{Effects of splitting}\label{justify:splitting}

Because we split our data and use the two halves in our estimator, the quantity
we are estimating actually corresponds to the FVR for a data set with half as
many observations.  As in cross validation, it is reasonable to wonder what
effect this has on our estimate.

The sample size influences the true FVR only through differences in the
resulting variable orderings from stepwise selection.  As a result, we expect
the true FVR values for both sample sizes to be reasonably close, and for the
uncertainty in FVR estimation to dominate the difference in most cases.  This
is supported by the simulations of Section \ref{simulation}.  

In the cases
where the true FVR values do 
diverge, the value for the half-sized data set will be larger, leading our
estimate to be conservative.  We observe this for particularly noisy data in
Figure \ref{estsim:bad}.

\subsection{Simulation}\label{simulation}

Here we present simulations of the performance of our estimation method for FVR
in
stepwise regression.  We will see that the method performs well overall.  We
will also simulate under parameters specifically selected to demonstrate the potential
biases discussed in Section \ref{justify}.

For these simulations, we use blocked settings similar to those used in Section
\ref{understanding:simulation} and illustrated in Figure \ref{img:matrixpic}.  These examples are constructed to clearly
illustrate the difference between FVR and FDR, and to show how our estimates
relate to those quantities.  To do so, the parameters are chosen so that the
blocks of variables will be reasonably significant, but the variables within
the blocks are correlated enough to be difficult to distinguish in the presence
of noise.  For convenience, the parameters of all the
simulations are laid out in Table \ref{simparamtable}.  All of these
simulations use $\lambda=0.5$ for the threshold in the estimator.  All
estimates are obtained by averaging 50 splits of the data set.  All curves and
standard errors are the result of 100 Monte Carlo simulations.

\begin{table}[tbh]
  \centering
  \begin{tabular}{r|c|c|c|c|c|c|}
& $n$ & $\#$ blocks & $\#$ per block & $\#$ signal & $\sigma_{\eps}$ &
$\rho$\\
    \hline
    Figure \ref{estsim:nice} & 100 & 20 & 2 & 6 & 0.8 & 0.95 \\
    Figure \ref{estsim:low} & 100 & 5 & 3 & 3 & 0.5 & 0.95\\
    Figure \ref{estsim:high} & 100 & 20 & 2 & 10 & 0.5 & 0.95\\
    Figure \ref{estsim:bad} & 100 & 20 & 2 & 10 & 2 & 0.95\\
    \hline
  \end{tabular}
  \caption{Parameters for the simulation settings use to make Figures \ref{estsim:nice}, \ref{estsim:low},
  \ref{estsim:high} and \ref{estsim:bad}.  The parameters are number of
  observations, number of blocks of variables, number of variables per block,
  number of blocks where one variable is made significant, noise variance, and
  within block correlation.
  The coefficient for any significant variables is fixed at 1 across all
  simulations, and all estimators use $\lambda = 0.5$.}
  \label{simparamtable}
\end{table}

\begin{figure}[h!]
  \centering
  \includegraphics[width=0.7\linewidth]{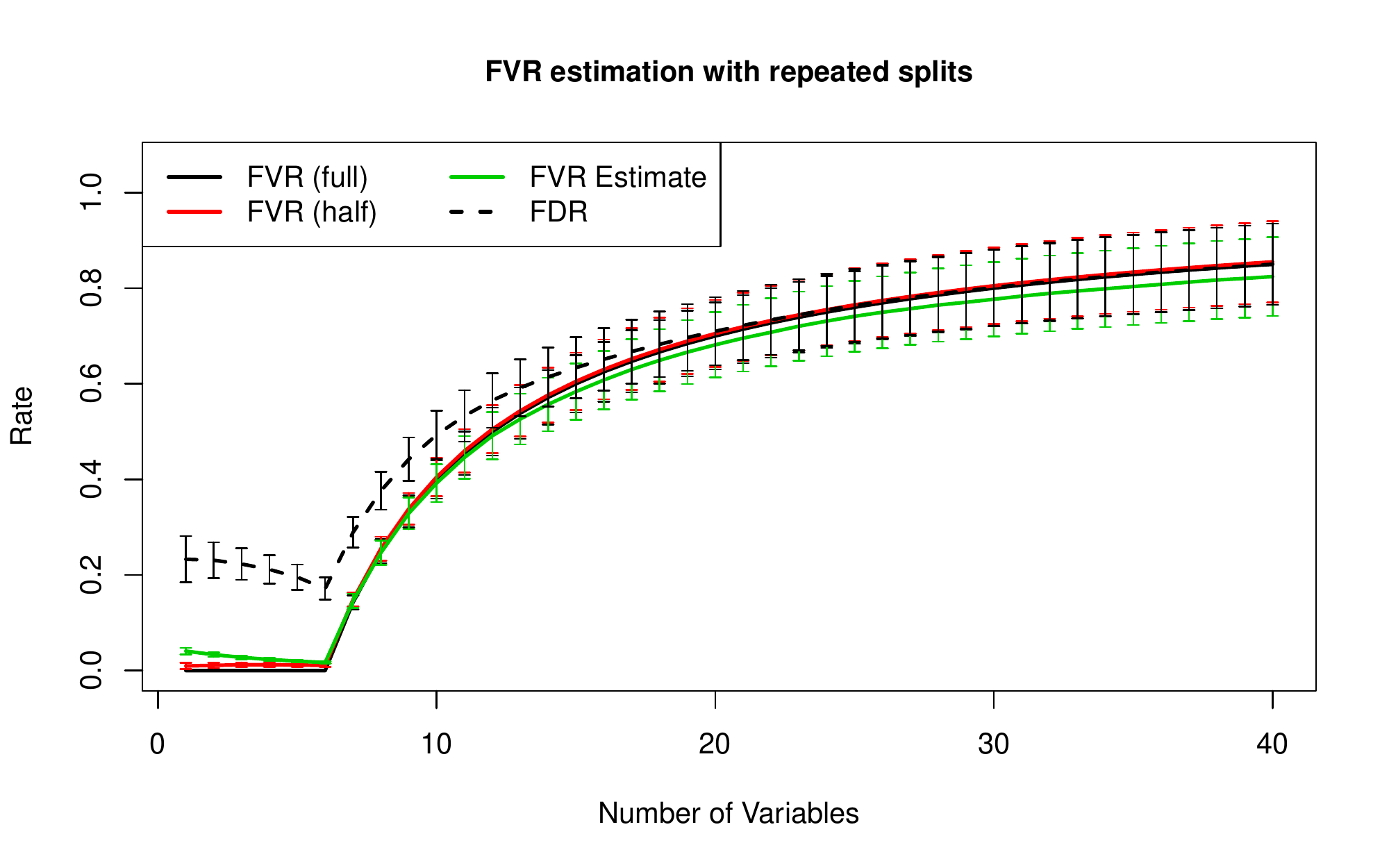}
  \caption{Simulation of the FVR estimate for forward stepwise selection on
  blocked data in an ideal setting.  The true full model FDR is shown in black,
  while the true FVR is shown in black for selection with the full sample size,
  and red for selection with a half sample size.  The estimated FVR, shown in
  green, closely follows both of these true FVR curves.}
  \label{estsim:nice}
\end{figure}

Ideal performance of our estimate can be seen in Figure \ref{estsim:nice}.
The plot shows the true FVR for both the full
sample size and the half sample size (in black and red, respectively), along with the true FDR (dotted black) and our estimate (green).
We show both of
these true FVR values to demonstrate that splitting our sample has not
dramatically altered our target quantity, as discussed in Section
\ref{justify:splitting}.

In the simulation shown in Figure \ref{estsim:nice}, stepwise selection
correctly selects the six groups with signal, but not the particular variable
within each of those groups.  Thus the FDR is reasonably large from the
beginning, as we would expect from the construction of the example, while the
FVR remains low.  We see that the FVR estimate, using the method from Section
\ref{estimation:algorithm}, closely matches the true FVR for both sample sizes.

\begin{figure}[h!]
  \centering
  \includegraphics[width=0.7\linewidth]{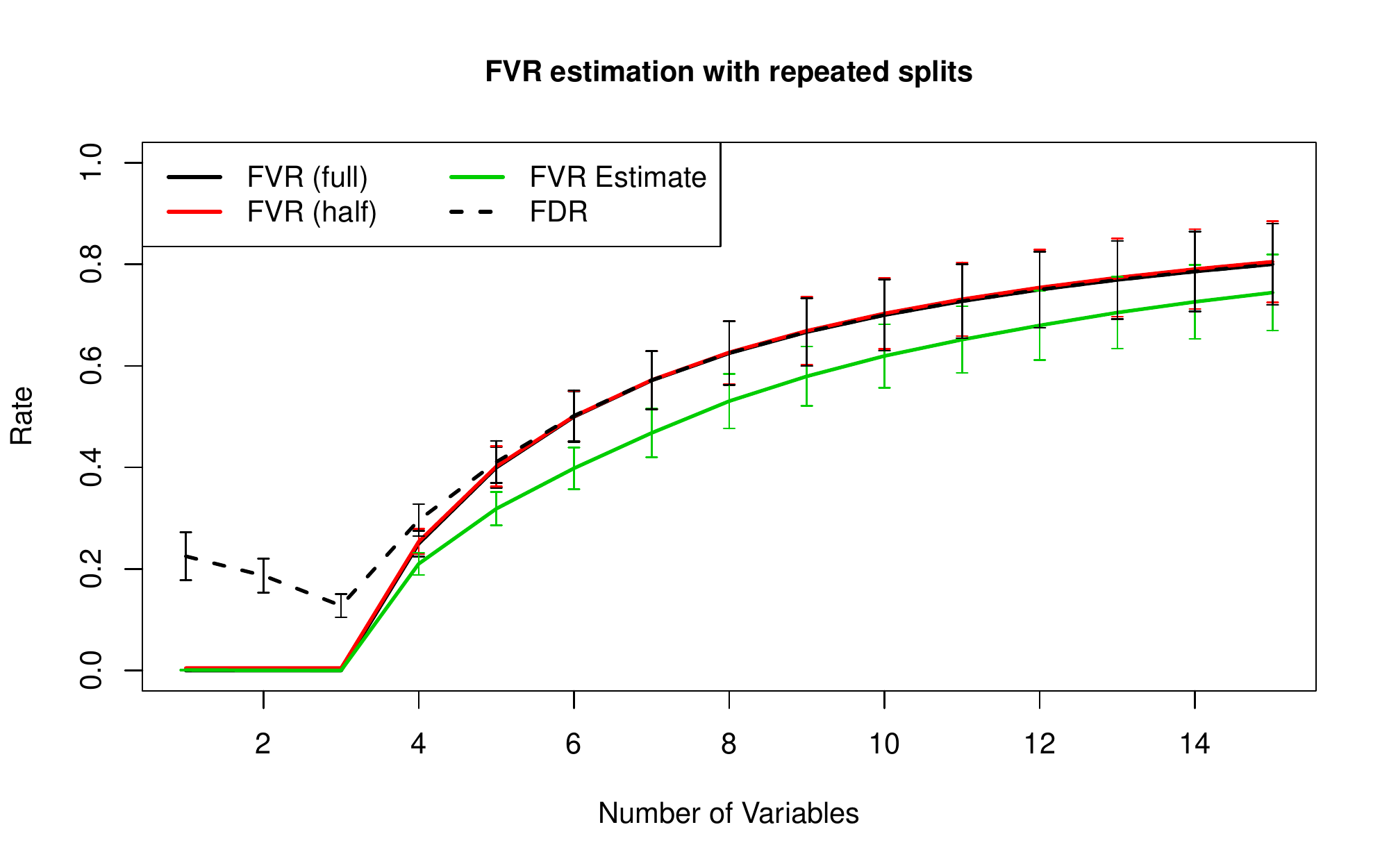}
  \caption{Simulation of the FVR estimate for forward stepwise selection on
  blocked data, demonstrating downward bias due to misordering of selected
  variables.  The right side of the path shows a downward bias in the
  estimates (green) relative to the true FVR values (black and red).}
  \label{estsim:low}
\end{figure}

The simulation in Figure \ref{estsim:low} illustrates the low bias in our
estimator due to incorrect ordering of the variables, as discussed in Section
\ref{justify:ordering}.  This simulation is constructed to have a few signal
variables, along with many very strongly correlated noise variables in
relatively low noise.  When the noise variables enter early in the path, they
temporarily appear informative due to spurious correlations.  This causes a
downward bias in the estimated number of false selections.

\begin{figure}[h!]
  \centering
  \includegraphics[width=0.7\linewidth]{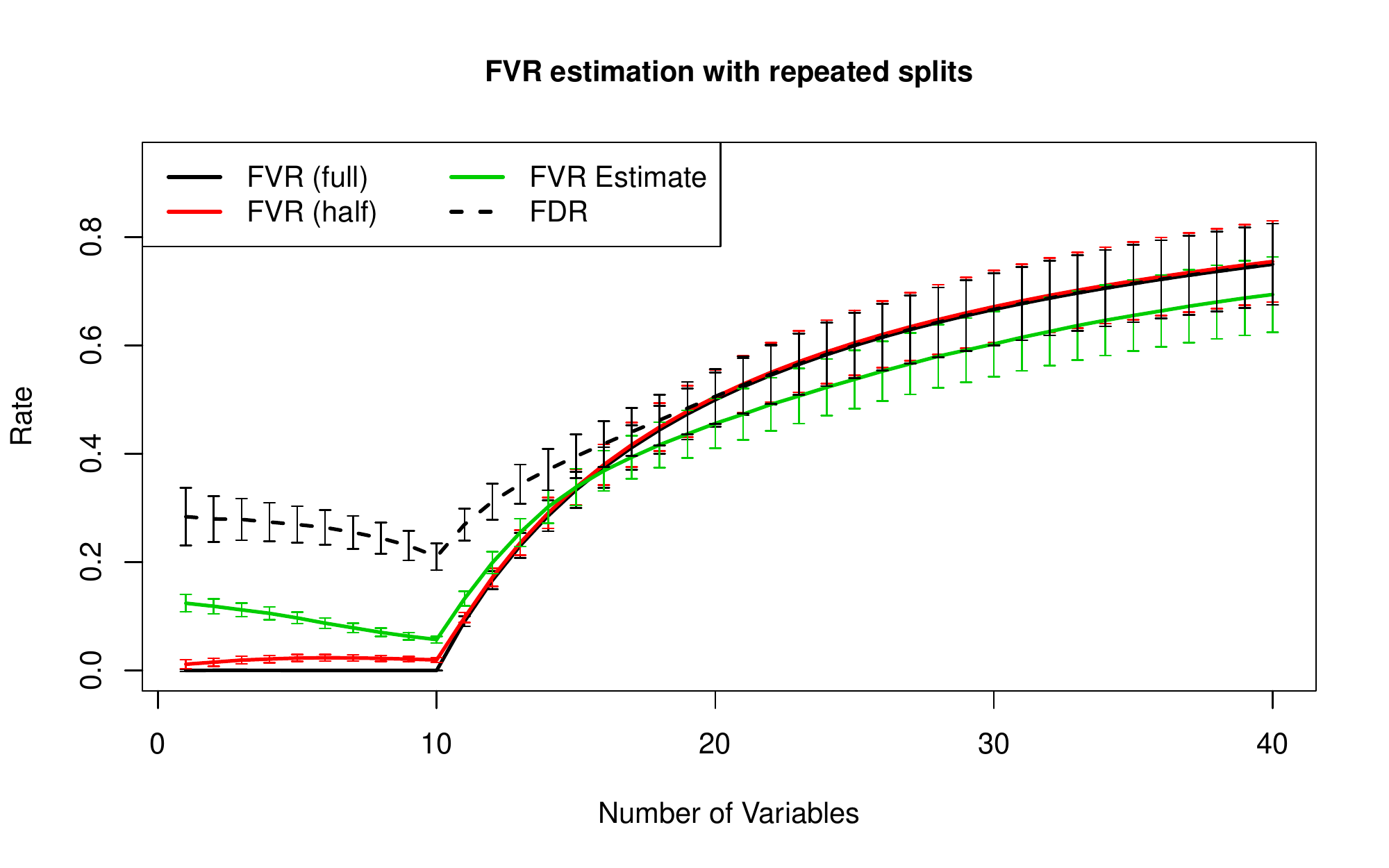}
  \caption{Simulation of FVR estimate for forward stepwise selection on blocked
  data, demonstrating the upward bias
  of the threshold estimator.  The left side of the path shows an upward bias
  in the estimates (green) relative to the true FVR values (red and black).}
  \label{estsim:high}
\end{figure}

The simulation in Figure \ref{estsim:high} illustrates the upward bias in
the threshold estimator that was discussed in Section \ref{justify:storey}.
Here many signal signal variables into
the simulation.  The FVR estimate (green) is
biased upward from the true values (black/red), particularly at the start of
the path.  While this is worrisome, it is comforting that the dramatic bias is
upward, leading to conservative estimates.  The selection of $\lambda$ in this
simulation was made without tuning to reduce this bias; bootstrap calibration
as in \cite{storey2002} might help to reduce this bias.

\begin{figure}[h!]
  \centering
  \includegraphics[width=0.7\linewidth]{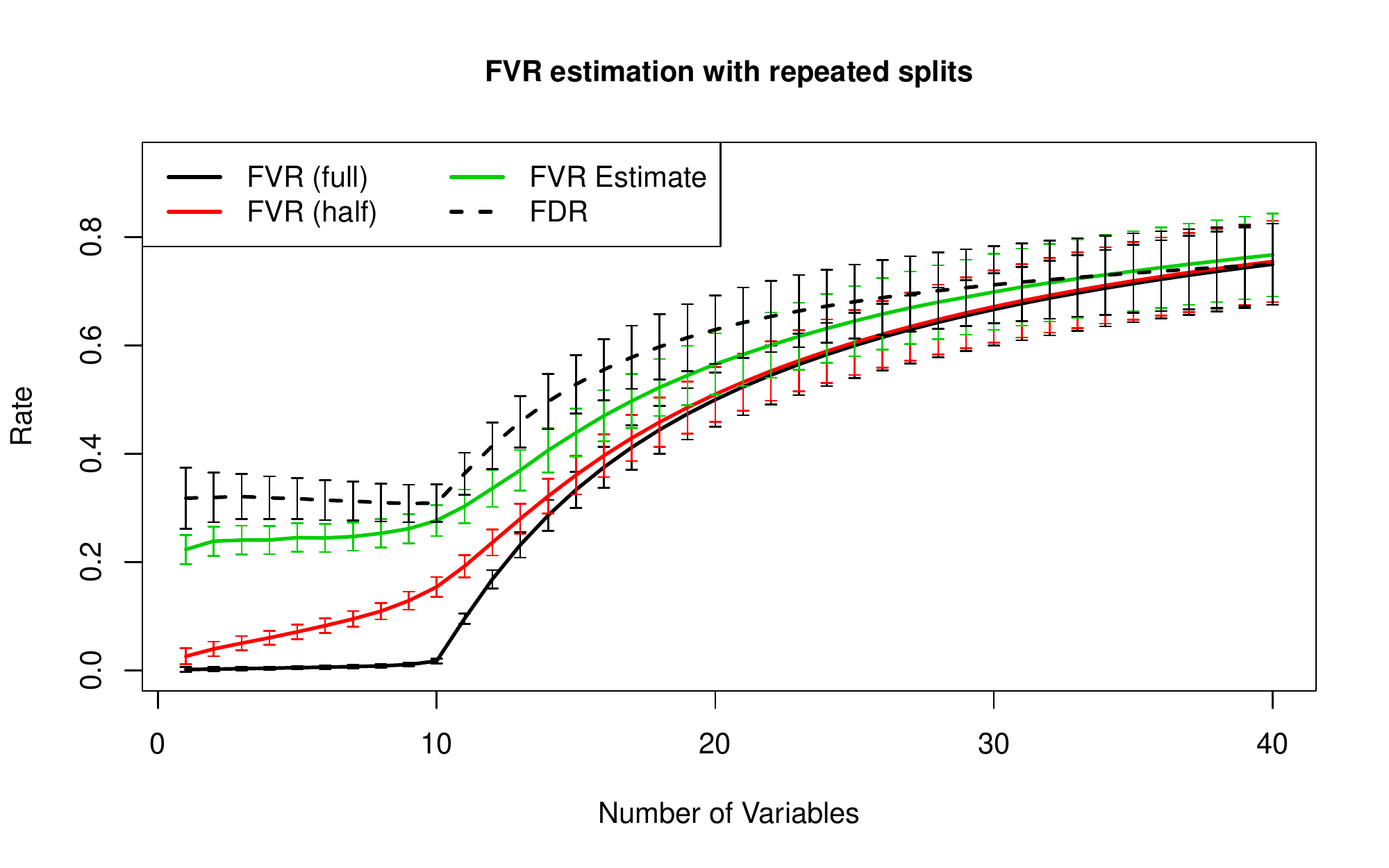}
  \caption{Simulation of FVR estimate for forward stepwise selection on blocked
  data in a high-noise setting.  This setting shows that, in the presence of
  high noise, the lack of data in the half samples cause that true FVR (red) to
  be larger than the true FVR for a full sample (black).  This inflates
  our estimator, since our estimator splits the data and actually estimates the
  half-sample FVR.}
  \label{estsim:bad}
\end{figure}

Finally, the simulation is conducted with a much higher noise level, shown in
Figure \ref{estsim:bad}. This has the effect of inflating the
difference between true FVR values for the full and half samples.  The half sample FVR (red) is now larger than the
full sample FVR (black), implying that the additional information from
the larger sample is important for obtaining good selections.  Our estimate
(green) is estimating the higher of these curves, and is therefore very
conservative.  

We see from these simulations that the proposed estimator works reasonably well
in simulation.  The biases discussed in Section \ref{justify} do exist.  The
downward bias from relying on the stepwise selection ordering appears to be
weak in practice and to occur mostly in the later part of the selection path,
supporting our belief that stepwise selection is providing a reliable
ordering.  The upward bias from the threshold estimator of true null hypotheses
appears when there are many signal variables present in the data, but it skews
the results in a conservative direction.

\section{Conclusion}

In this paper, we discussed the interpretations and implications of different
definitions of false selection in the regression setting.  We saw that these
error criteria behave differently in cases where variables are correlated.  In
particular, we described difficulties for the standard full model definition,
which lead to unintuitive or undesirable behavior in many cases.  

As a solution, we introduced a new false selection error criteria, FVR, which
is defined in terms of the projected model.  This error criterion focuses on
guaranteeing uniqueness of the explanatory variables only among the selected
variables, rather than the entire data set.  In doing so, it avoids
the concerning behaviors of the traditional full model definition, leading to intuitive
behavior in many settings.  We presented several interpretations of FVR,
demonstrating its differences from FDR and where each criterion might be appropriate to
use.

Finally, we presented a simple estimation method for FVR in stepwise
regression.  We showed that this method gave reasonable estimates of FVR over a
range of simulation parameters.  We also examined the regimes in which this
estimator performed poorly, giving insight that could be helpful when
constructing future estimators or assessing existing ones.

The idea that each of the error criteria impose a different idea of an
``interesting'' selected variable is a convenient view.  The full model
definition considers a selected variable interesting if it explains signal that
is not captured by any other variable in the data set.  In contrast, the
projected model definition (corresponding to FVR) considers a selected variable
interesting if it captures unique signal only among the other variables in the
data set.  Contrasting the error criteria in this way could help provide an
intuition of which criterion is most suitable to a particular problem.

There is plenty of interesting work to be done in understanding this new error
criterion.  More work is needed to understand better estimation or control
procedures.  Very preliminary work suggests that other common variable
selection procedures like the LASSO \citep{roblasso} may control FVR well.  Existing
methods that seek to control FDR in the regression setting are also likely to have
appealing FVR properties.  In another direction, one can construct an analog to
the False Negative Rate of \cite{fnr} which is related to FVR and might be
interesting to consider.  Our main goal in this paper has been to identify potential concerns with the
accepted full model FDR definition when predictors are correlated, and to
propose this new FVR criterion which may be more appropriate to consider in those settings.

\appendix
\section{Details}\label{app:details}
\subsection{Gaussian formulation of $\mathrm{FVP}$ in terms of the covariance
matrices}\label{app:details:gaussian}

It is particularly straightforward to compute the FVP in the joint multivariate
Gaussian setting where the parameters are known.  This is particularly useful
when running simulations, so we include our approach here.

Suppose that $(x^{(1)},\dots,x^{(p)},y)^T$ is joint multivariate normal, with
$X\sim N(0,\Sigma)$ and $y\sim N(X\beta,\sigma_{\eps}^2)$.  Note that
\begin{align*}
  \Cov(y_i,x_i^{(j)}) &= \E(x_i^{(j)} y_i) = \sum_{j'=1}^p\beta_{j'}
  \E(x_i^{(j)}x_i^{(j')}) = \sum_{j'=1}^p
  \beta_{j'} \Sigma_{jj'}\\
  \Cov(y_i,y_i) &= \Cov(x_i^T\beta,x_i^T\beta) + \sigma_{\eps}^2 =
  \E\left(\left(\sum_{j=1}^p x_i^{(j)}\beta_j\right)\left(\sum_{j'=1}^p
  x_i^{(j')}\beta_{j'}\right)\right) + \sigma_{\eps}^2\\
  &= \sum_{jj'} \beta_j\beta_{j'} \E(x_i^{(j)}x_i^{(j')}) = \beta^T\Sigma\beta
\end{align*}

This lets us construct the augmented covariance matrix for
$(x^{(1)},\dots,x^{(p)},y)^T$, $\tilde{\Sigma}$.
\begin{align*}
  \tilde{\Sigma} = \begin{pmatrix}  
    \Sigma & \Sigma \beta \\
    \beta^T \Sigma & \beta^T\Sigma\beta + \sigma_{\eps}^2
  \end{pmatrix} 
\end{align*}

Now suppose we select a set of variables $\Ac$.  We want to assess the FVP of
this set of variables.  Let $\Ac^+$ be $A \cup y$.  The covariance matrix for
the marginal distribution on $\Ac^+$ is just $\tilde{\Sigma}_{\Ac^+,\Ac^+}$.
We can compute the inverse $\tilde{\Sigma}_{\Ac^+,\Ac^+}^{-1}$, and the FVP is
the number of zeros in the row corresponding to $y$.

\subsection{Selecting $\lambda$ for the threshold estimator}\label{app:justify:threshold}

In Section \ref{estimation:algorithm}, we introduce a threshold estimator
$\hat{V}_k^{(\lambda)} = \frac{\#\{p_j > \lambda; j\le k\}}{(1-\lambda)}$
for estimating the number of true nulls in our set of hypotheses, and show that 
the expectation of this estimator provides a lower bound on
the number of true nulls in $\{H_j^{(\Pc)}i; j\le k\}$.

As mentioned in Section \ref{justify:storey}, there is a bias-variance
trade-off involved in selecting $\lambda$.  As $\lambda$ increases, the
probability of a false null hypothesis entering will decrease, decreasing the
bias in the estimator.  However, the probability that a true null hypothesis is
counted will also decrease, increasing the variance of the estimator.  

In \cite{storey2002}, a bootstrap-based approach is presented for tuning
$\lambda$ in a similar threshold estimator of $\widehat{pFDR}$.  It can be shown that an
equivalent condition, $\E\hat{V}_k^{(\lambda)} \ge \min_{\lambda'}\E
\hat{V}_k^{(\lambda)} \ge V_k$, holds for our estimator, so a similar tuning
approach could be used to select $\lambda$ for a particular
application.  This would mean choosing $\lambda$ to minimize 
\begin{align*}
  \frac{1}{B} \sum_{b=1}^B
\left(\hat{V}_k^{(\lambda)\star b}- \min_{\lambda'}
\hat{V}_k^{(\lambda')}\right)^2,
\end{align*}
where $b$ indexes $B$ bootstrap simulations,
each of which contributes a bootstrap estimate $\hat{V}_k^{(\lambda)\star b}$.

\subsection{Effects of variable ordering}\label{app:justify}

Here we present more details in our discussion of variable ordering and the
estimator of Section \ref{estimation}.

Let $\Ac$ be the selected set of variables and $V$ be the number of variables
in $\Ac$ with zero coefficients in the model projected onto $X_{(\Ac)}$, making
$V$ the numerator of the FVP for $\Ac$.  Let $\Bc \subseteq \Ac$ be a minimal subset of the variables in $\Ac$, such
that the projection of the true model $X\beta$ onto $X_{\Bc}$ is the same as
the projection of that 
model onto $X_{\Ac}$.  Note that $V = |\Ac|-|\Bc|$.

Suppose that $\tilde{\Pc}$ is an ordering of the variables in $\Ac$ such that all the
variables in $\Bc$ appear before any other variables.  Then, except for very
special cases, the incremental null hypothesis is false at each of the first
$|\Bc|$ steps along $\tilde{\Pc}$.  Furthermore, the subsequent $|\Ac|-|\Bc|$
steps are true null hypotheses, since the projected model has been obtained
once the variables in $\Bc$ have been included.  This means that there will be
exactly $V = |\Ac|-|\Bc|$ true null incremental hypotheses in $\{H_j^{(\tilde{\Pc})}\}$,
which is identical to the number of zero coefficients in the projected model.

The special cases above refer to the unlikely cases where some of the variables
in $\Bc$ have exactly zero correlation with $y$ when a strict subset of $\Bc$
are conditioned upon.  However, while the statements above will not hold in
that case for all orderings where $\Bc$ appears first, there will still exist a
subset of such orderings for which the statements hold.  Furthermore, stepwise
selection will tend not to select the invalid orderings, so the issue should
not arise in practice.

In practice, an ideal path $\tilde{P}$ is not known.  Our algorithm relies on
the idea that the paths produced by stepwise selection are close to one of
these perfectly-ordered paths.  We can view a real path $\Pc$ as a modification
of a ``nearest'' path $\tilde{P}$, where the path is modified by moving noise
variables forward in the ordering, and the path $\tilde{P}$ is nearest if it
requires the fewest such moves.  If $t$ is the number of these erroneous moves,
then at worst all $t$ improperly inserted variables will appear significant at
the time of their selection.  In that worst case, the quantity being estimated by looking at the
incremental null hypotheses is actually $\mathrm{FVP} - t/\Ac$.  

This means that an improper ordering could bias the FVR estimate low, by as
much as $t/\Ac$.  The fact that this bias is downward is worrisome, since
it could potentially lead to over-optimistic estimates of FVR.  In simulation, we have found that stepwise regression
produces orderings that are quite reasonable, leading to small $t$
and minimal bias.  It could be interesting to investigate the settings in which
$t$ can be shown to be controlled by stepwise selection or other methods.

\section*{Acknowledgements}
The authors would like to thank Noah Simon, Alexandra Chouldechova, and Jacob
Bien for helpful discussions.

\bibliographystyle{imsart-nameyear}
\bibliography{FVRpaperbib}

\end{document}